\documentclass[%
 preprint,
 amsmath,amssymb,
 aps,
]{revtex4-1}

\usepackage{graphicx}
\usepackage{color}
\usepackage{dcolumn}
\usepackage{bm}
\usepackage{natbib}
\usepackage{footnote}
\newcommand{\average}[1]{\ensuremath{\langle#1\rangle} }
\begin{document}
\preprint{APS/123-QED}

\title{Financial Knudsen number: breakdown of continuous price dynamics and
asymmetric buy and sell structures confirmed by high precision order book information
}

\author{Yoshihiro Yura$^{1}$, Hideki Takayasu$^{2,3}$, Didier Sornette$^{4}$ and Misako Takayasu$^{1}$$^*$}
\affiliation{
$^{1}$Department of Computational Intelligence and Systems Science, Interdisciplinary Graduate School of Science and Engineering, Tokyo Institute of Technology 4259 Nagatsuta-cho, Yokohama 226-8502, Japan.
}
\affiliation{%
$^{2}$Sony Computer Science Laboratories, 3-14-13, Higashi-Gotanda, Shinagawa-ku, Tokyo, 141-0022, Japan.
}
\affiliation{
$^{3}$Meiji Institute for Advanced Study of Mathematical Sciences, Meiji University, 
4-21-1 Nakano, Nakano-ku, Tokyo, 164-8525, Japan.
}
\affiliation{
$^{4}$ETH Zurich, D-MTEC, Scheuchzerstrasse 7, 8092 Zurich, Switzerland.
}

\begin{abstract}
We generalise the description of the dynamics of the order book of financial markets 
in terms of a Brownian particle embedded in a fluid of incoming, exiting and annihilating particles 
by presenting a model of the velocity on each side (buy and sell) independently. The improved model
builds on the time-averaged number of particles in the inner layer and 
its change per unit time, where the inner layer is revealed
by the correlations between price velocity and change in the number of particles (limit orders).
This allows us to introduce the Knudsen number of the financial Brownian particle motion
and its asymmetric version (on the buy and sell sides). Not being considered previously,
the asymmetric Knudsen numbers are crucial in finance in order to detect asymmetric price changes.
The Knudsen numbers allows us to characterise
the conditions for the market dynamics to be correctly described by a continuous stochastic
process. Not questioned until now for large liquid markets such as the USD/JPY and EUR/USD exchange rates,
we show that there are regimes when the Knudsen numbers are so high that 
discrete particle effects dominate, such as during market stresses and crashes.
We document the presence of imbalances of particles depletion rates on the buy and sell sides
that are associated with high Knudsen numbers and
violent directional price changes. This indicator can
detect the direction of the price motion at the early stage while the usual volatility risk measure
is blind to the price direction. 
\end{abstract}

\maketitle

\section{Introduction}

The mean free path length, 
the average distance travelled by a moving particle between successive collisions, 
is an important quantity to characterise fluids~\citep{Chapman}. 
In statistical physics, especially hydrodynamics, the Knudsen number~\citep{Knudsen,Fluid,PhysRevE.74.031402}, 
which is defined as the ratio of the mean free path length to a representative particle length scale, 
 is used to determine whether the system can be described in the continuum limit or 
 needs a description accounting for discrete or particle effects. 
Generally, the fluid can be well approximated by continuous mathematical equations
when the Knudsen number is below $0.1$, so that sufficiently many collisions 
occur within a particle size to erase the influence of any discreteness. Thus, for 
a Knudsen number below $0.1$, one uses in general the Navier-Stokes equation,
while the Boltzmann equation is needed for Knudsen number larger than $1$, with 
a complicated transition in between. There are long standing problems characterised
by high Knudsen numbers, including dust particle motions through the lower atmosphere, 
satellite motion through the exosphere, microfluidics \citep{Niimi} and micromechanical systems.
These ``high Knudsen number flows'' possess novel properties not shared
by fluid flows at low Knudsen numbers, which are being investigated with novel 
instruments and theoretical methods ~\citep{Oh,Kumaran,William,PhysRevE.82.026307}.

Beyond the scale of the mean free path, particles follow random paths, a large
class of which are called Brownian motions. This later concept describes the random motion of 
small objects in fluid media, which was discovered by 
Jan Ingenhousz in 1784 when observing the irregular motion of coal dust 
particles on the surface of alcohol~\citep{PWvanderPas} and 
Botanist R. Brown in 1827 when observing pollen grains under a microscope. 
The mechanism of this phenomenon was finally clarified 
by independently Einstein and Smoluchowski, explaining
the random motion of the Brownian particle as due to the incessant collisions
with the embedding fluid particles, proving theoretically the particle nature of matter \citep{Einstein,Smoluchowski}. 
This idea is one of the fundamental starting points of statistical physics, which 
connects micro-scale to macro-scale phenomena.

Several years before Einstein's random walk theory, 
Bachelier introduced the random walk model to describe the dynamics 
of prices of assets traded in stock markets \citep{Bachelier1900}.
Indeed, prices of traded assets fluctuate incessantly and randomly up and down
as a result of the aggregation of all traders' decisions. The concept of 
random walks to characterise the aggregate actions of motivated and diligent investors
was a significant breakthrough that has further developed into the
``efficient market hypothesis'' (see \cite{SornetteRev2014} for a short history and review),
which forms the foundation of financial engineering~\citep{Merton}. 
Finance and physics are both founded on the theory of random walks
and their many generalisations, going back to more than one hundred years
of intertwined history \cite{SornetteRev2014}.
However, the meaning of the Knudsen number in finance has not
be elucidated until now and this will be the principal goal of this article.
 
In 1963, Mandelbrot documented that the scale of volatility 
(defined as the absolute price changes in a given 
time interval) is nonstationary and the probability distribution of price change
is asymptotically described by a power law tail \citep{Mandelbrot}. These two properties
are now recognised as fundamental stylised facts of financial price series. 
Building on analogies with the many power law distributions in physics, this
has later excited the interest of statistical physicists to analyze financial time series further, 
using the large amount of available data. 
Recently, several novel properties of financial prices series have been documented, such as
the negative auto-correlation of price changes at short times, the abnormal diffusion of prices~\citep{Takayasu_PUCK,Watanabe,Yura}, 
the long-auto correlation of volatility~\citep{Bollerslev86generalizedautoregressive}, 
the long-memory process of sign of orders~\citep{RePEc:bpj:sndecm:v:8:y:2004:i:3:n:1,Bouchaud4},
the property that implicitly underlies the present considerations as financial multifractality~\citep{2005626,2012115}, 
large price changes characterized by 
the gap of a limit order book (containing no quotes between prices)~\citep{doi:10.1142/S0219477505002574},
the existence of endogenous feedback mechanisms that are
well characterised by the self-excited Hawkes model~\citep{PhysRevE.85.056108}, 
rich market impact dynamics revealed in nonlinear price changes caused by submitted orders~\citep{PhysRevX.1.021006,PhysRevE.89.042805,PhysRevE.71.046131}, 
and the Fokker-Plank description for the queue dynamics of the order book~\citep{PhysRevE.88.032809}.

In this paper, we analyze financial markets from the micro scale view of traders~\citep{Abergel,Eisler,Huang}, 
We focus on a foreign exchange market, which is the place where 
a trader can exchange his currency to another currency via a trade with another trader. 
As an example, in the USD-JPY currency market, 
we call a trader who wants to change JPY to USD a buyer and USD to JPY a seller. 
There are two possible actions available to each trader. The first one is to submit a market order that 
is immediately executed at the best price in the market at the time of the submission.
The second possible action is a limit order that enables a trader to book his order in the market with a specific 
price and quantity, and which can be canceled later if he wishes. If two limit orders
have been placed at the same price, the tie is broken on a first-come first-served basis.
The aggregation of all limit orders is called the order book, in which the limit orders
both on the buyer side (bids) and on the seller side (asks) are placed on 
the price axis while the volume associated to each limit price is 
represented as a 2-D graph to be shown later. A trade occurs when either a market
order or a limit order meets the demand/supply of existing limit orders.
The prices and volumes in the order book and thus the realised market price time series
result from the aggregation of all traders' decision making processes. 
Describing in details the full dynamics of the order book process is a challenging task
that can benefit from the perspective of statistical physics applied to financial markets.

Recently, the celebrated fluctuation-dissipation Theorem (FDT), which states that 
the response of a system in equilibrium to a small applied force is the same as its response 
to a spontaneous fluctuation \citep{Kubo,GLE}, has been found to hold 
for the average dynamics of the limit order prices in the order book \citep{PhysRevLett.112.098703},
suggesting a novel approach to better understand the traders' decision making process.
One can indeed observe the traders' immediate reactions to price changes 
in terms of the placement of their orders in the order book. The remarkable finding
is that the traders' orders, which may be realised in the future through a trade, act similarly to fluid particles
ahead or behind a Brownian particle (whose position is the mid-price defined as the average
of the best bid and ask prices) in the physical fluid analogy. Hence the FDT
expresses a remarkable relationship between the fluctuations of the mid-price
and an effective drag force exerted by limit orders at different locations in the order book.
These properties were found to be well described by a Langevin equation. 

These results provide a bridge between the conceptual foundations 
of the random walk theory and its generalisation to both Brownian particles in physics
and stock market prices in finance. The random walk model of financial prices
has been since Bachelier thought to be the consequence of the no-arbitrage condition (called the Martingale
property in mathematical finance) and the observable embodiement
of the efficient market hypothesis. In contrast, our model \citep{PhysRevLett.112.098703} in terms of a
financial Brownian particle in the layered order book fluid and the existence
of the fluctuation-dissipation theorem provides a novel picture in which the
random price walks are, like pollen particles in a fluid, the result of the ``collisions''
with the limit orders that are continuously jiggling, appearing and disappearing.

Unravelling the physical underpinning of fluctuation phenomena
in term of the underlying random walk behaviors provides both intuitive and deeper
understanding in a variety of systems \citep{sornette,PhysRevLett.53.633}. Here, we use this
novel understanding of financial price dynamics to investigate the conditions under which the standard
mathematical description \citep{Hull} of financial price dynamics using Wiener process
(the continuous limit of random walks) and stochastic Ito calculus hold. Informed
by the underlying discrete nature of the limit orders acting as effective embedding
fluid particles, we introduce a financial Knudsen number, with the purpose
of testing the validity of the continuous mathematical models of finance \cite{Jeanblancetal2009}.
In section 2, we present the data analysis of the foreign exchange data set. 
In section 3, we introduce the Knudsen number of the market.
In section 4, we relate the Knudsen number observed in the historical time series to 
the periods when the market price exhibits ballistic-like motions.  Section 5 concludes.

\section{Preliminary analysis of the dynamics of limit order books}

We have analyzed the databse of the Spot FX markets data provided by EBS (ICAP), 
which is known as one of the biggest FX markets mainly for inter-bank dealers. 
It includes the order book information of 
USD/JPY, EUR/USD, and EUR/JPY during the week beginning on March 14, 2011.
In the data, every order is recorded every millisecond, and a minimum price unit  
is fixed at 0.001[YEN] for USD/JPY and EUR/JPY, 
0.00001[USD] for EUR/USD. 
In the following, the minimum price unit is represented by $\Delta x$
(thus $\Delta x=0.001$[JPY] for USD/JPY and EUR/JPY, $\Delta x=0.00001$[USD] for EUR/USD.)

To analyze the financial time series, we use an event time rather than calendar time. 
An event time is counted after a transaction occurs. 
After the $k$-th transaction, the event time $t$ is thus $k\Delta t ~(k=1,2,...)$, 
where $\Delta t$ is the average waiting time between two successive transactions. 

\subsection{Definition of variables for data analysis}

Let us now present the definitions of the relevant variables. 
At time $t(=k\Delta t)$, 
the number of limit orders on the price axis $x$ is given by $N(x,t)(\ge 0)$.
The minimum value and unit of $N(x)$ is 1 million dollars in USD/JPY, 
denoted by $\Delta n$ 
(1 million euro in EUR/JPY and EUR/USD). 

We represent the best price on the positive ($+$) side, that is the lowest sell price in the order book, as $x^{+}(t)$, and the best price on the negative ($-$) side, as $x^{-}(t)$ for the highest buy price, 
and the market price is defined as the mid-price
\begin{eqnarray}
x(t)\equiv \{x^{+}(t)+x^{-}(t)\}/2. 
\end{eqnarray}
In the time interval 
[$t-\Delta t,t$], 
the velocity of the price (i.e. price change per unit time) is defined as 
\begin{eqnarray}
v(t)\equiv \{x(t)-x(t-\Delta t)\}/\Delta t. 
\end{eqnarray}

The depth of the market is introduced to describe the position of limit orders against the best price on its side. 
At time $t$, on the $j(=+/-)$ side, the position $x$ of a limit order is represented as $Q^j(t)$. 
The depth $\gamma(Q^j,t)$ of the market of a specific limit order with respect to the best price at time $t-\Delta t$ is defined as 
\begin{eqnarray}
\gamma(Q^j,t)& \equiv & {\rm sign}(j) \left( \frac{Q^j(t) - x^j(t-\Delta t) }{ \Delta x }\right), 
\label{R_g}
\end{eqnarray}
where the depth is counted as multiples of a minimum unit $\Delta x$, and sign$(+)=1$ and sign$(-)=-1$. 
In cases when there is no need to show the position $x$ and time $t$, $\gamma(Q^j,t)$ is simply described as $\gamma$, 
When we need to specify the side $j$, $\gamma$ is shown as $\gamma^j$. 
The sign in the $-$ side in Eq.(\ref{R_g}) is needed to make positive the depth of the market. 
The number of limit orders at depth $\gamma$ is represented by $N_{\gamma}(\gamma,t)$. 
Next, the cumulative volume of limit orders from depth $0$ to $\gamma$ along the market axis depth is defined as 
\begin{eqnarray}
V_{\gamma}(\gamma,t)\equiv \sum_{\gamma'=0}^{\gamma} N_{\gamma}(\gamma',t).\label{Cum_V}
\end{eqnarray}

Finally, we consider the changes of configuration of limit orders at depth $\gamma$ along the market axis.
In the time interval 
[$t-\Delta t,t$], 
the number of limit order changes on the $j$ side at position $Q^j$
is defined as
\begin{eqnarray}
\Delta N_{\gamma}(\gamma(Q^j,t),t) &\equiv& N(Q^j,t) - N(Q^j,t-\Delta t) \label{Stationary1}~,
\end{eqnarray}
where the $\gamma$ dependence of $\Delta N_{\gamma}$ in the l.h.s. is obtained from
relationship (\ref{R_g}) between $\gamma$ and $Q^j$. 
Similarly, the change of the cumulative number of limit orders up to $\gamma$ is defined by
\begin{eqnarray}
\Delta V_{\gamma}(\gamma(Q^j,t),t) \equiv \sum_{\gamma'=0}^{\gamma} \Delta N_{\gamma}(\gamma',t)~.
\label{yjuytkikiotry}
\end{eqnarray}

\begin{figure}
\includegraphics[scale=1.2]{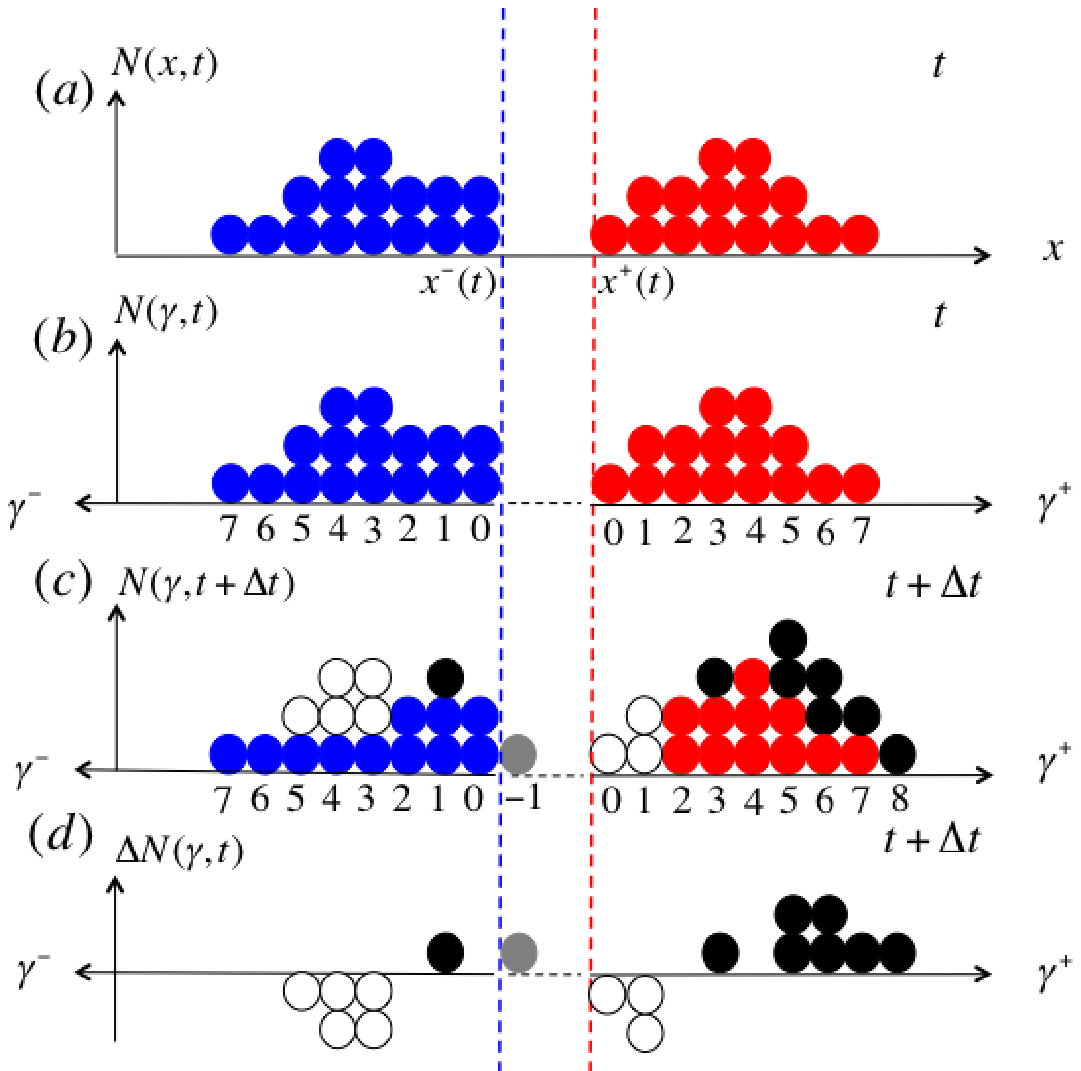}
\caption{
Schematic representation and definition of variables encoding the information contained in the order book.
(a) A configuration of the order book at time $t$ is shown on the horizontal price axis $x$ and the vertical
axis $N(x,t)$.  Blue circles correspond to limit orders on the $-$ side and red circles correspond 
to limit orders on the $+$ side. 
The dotted line coloured in blue indicates the best price on the $-$ side and 
red indicates the best price on the $+$ side. 
(b) The configuration in (a) is represented on the horizontal $\gamma^j$ axis and vertical $N_{\gamma}(\gamma,t)$ axis. 
Under the horizontal axis, the numbers indicate the depth values. 
(c) A configuration of the order book at time $t+\Delta t$ is shown on the horizontal $\gamma^j$ 
and vertical axis $N_{\gamma}(\gamma,t+\Delta t)$. 
Newly injected orders are coloured in black for $\gamma\ge0$ or grey for $\gamma<0$, while
canceled or executed orders are white. 
(d) The difference between the number of limit orders at $t+\Delta t$ and $t$ is shown 
on the horizontal $\gamma^j$ axis and vertical $N_{\gamma}(\gamma,t)$ axis. 
}
\label{fig1}
\end{figure}

Fig.\ref{fig1} explains the variables defined above. 
In Fig.\ref{fig1}(a), at time $t(=k\Delta t)$, 
a configuration of the order book is pictured along the horizontal axis $x$ and the vertical axis $N(x,t)$.
Blue and red circles represent limit orders in the $-$ side and $+$ side. 
The blue and red dotted line represents the limit orders boundary on the $-$ side and $+$ side, 
that is, 
$x^-(t)$ and $x^+(t)$, at time $t$. 
In Fig.\ref{fig1}(b), a configuration of the order book is pictured along the horizontal axis $\gamma$ 
and the vertical axis $N_{\gamma}(\gamma,t)$. Thus, Fig.\ref{fig1}(b) is derived
from Fig.\ref{fig1}(a) by changing the horizontal axis from price $x$ to the depth $\gamma$ of the market. 
The numbers shown below the horizontal axis are the values of $\gamma$ at each position. 
In Fig.\ref{fig1}(c), a configuration at $t+\Delta t$ is shown 
on the horizontal axis $\gamma$ and the vertical axis $N_{\gamma}(\gamma,t+\Delta t)$. 
Black and grey circles 
represent the newly injected limit orders in the time interval [$t,t+\Delta t$]. 
White circles represent canceled or executed limit orders. 
Black and gray circles correspond to $\gamma\ge0$ and $\gamma<0$. 
In Fig.\ref{fig1}(d), the change of configuration of the order book between $t$ and $t+\Delta t$
is shown on the horizontal axis $\gamma$ and the vertical axis $\Delta N_{\gamma}(\gamma,t)$. 
The changes of limit orders between Fig.\ref{fig1}(b) and (c), which are shown along the horizontal axis $\gamma$,
are encoded in the circle colours at each depth $\gamma$. 

The grey circle at $\gamma=-1$ in Fig.\ref{fig1}(c),(d) requires some explanation.
This negative $\gamma$ corresponds to a price position $Q^-=x^-(t)+\Delta x$ at $t+\Delta t$
falling inside the previous interval defined by the best bid and best buy orders at time $t$. The appearance of this new order
at time $t+\Delta t$ corresponds to the relations $N(Q^-,t)=0$, $N(Q^-,t+\Delta t)=1$ and
$\Delta N_{\gamma}(\gamma(Q^-,t),t)=1$. 
While the depth of the new order $\gamma(Q^-,t)$ is given by $\gamma(Q^-,t)=-\{(x^-(t)+\Delta x)-x^-(t)\}/\Delta x=-1$ 
according to definition Eq.(\ref{R_g}), it is measured as $\gamma=0$ at time $t+\Delta t$.
Generally, when $\gamma(Q^j,t) < 0$,
a newly injected limit order is placed lower for $j=+$ and higher for $j=-$ 
than the previous best price
in the corresponding ask and bid regions. 

\subsection{Properties of the cumulative limit orders as a function of depth}

\begin{figure}
\includegraphics[scale=1.2]{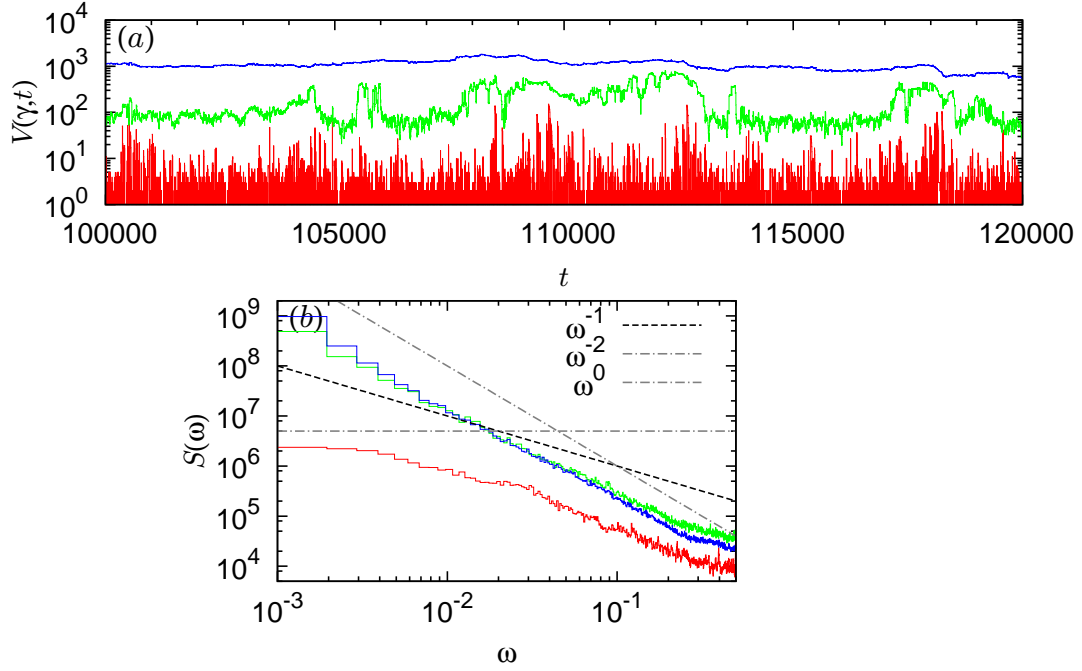}
\caption{
Study of the stationarity of the number of limit orders as a function of time:
(a) cumulative number $V_{\gamma}(\gamma,t)$ of limit orders up to depth $\gamma$
as a function of time $t$ for $\gamma=0$ (red), $\gamma=10^2$ (green) and $\gamma=10^3$ (blue).
(b) Power spectra $S_\gamma(\omega)$ of the three times series of panel (a) are shown 
as a function of angular frequency $\omega$, with the same colour code as in panel (a).
At long time scales ($\omega < 10^{-2}$), we can observe that $S_{\gamma=0}(\omega)$
becomes flat, i.e. with an exponent (defined in the text) converging to $0$, diagnosing
a stationary behavior. In contrast, both $S_{\gamma=10^2}(\omega)$ and $S_{\gamma=10^3}(\omega)$
have their exponent $\alpha \approx 2$, which is larger than $1$, thus expressing
a non-stationarity behavior of $V_{\gamma}(\gamma=10^{2},t)$ and $V_{\gamma}(\gamma=10^{3},t)$.}
\label{fig2}
\end{figure}

Let us investigate some properties of the cumulative limit orders up to depth $\gamma$, $V_{\gamma}(\gamma^j,t)$
defined by Eq.(\ref{Cum_V}), on both sides $j=+$ and $-$.
Fig.\ref{fig2}(a) shows the time evolution of $V_{\gamma}(\gamma^+,t)$ (where $t(=n\Delta t)$) for 
$\gamma=0, 10^2, 10^3$ (red, green, blue respectively). 
Although the time series for $\gamma=0$ has many pulse-like peaks, 
those for $\gamma=10^2$ and $10^3$ seem to behave like random walkers. 
To judge the stationarity of these time series, the power spectra of these time series are investigated. 
In Fig.\ref{fig2}(b), the power spectra are shown in log-log scale (with the
same colour code as in Fig.\ref{fig2}(a)). Approximating a given spectrum
by the power law $S(\omega)\sim 1/\omega^{\alpha}$, recall that a value $\alpha$ larger than $1$
in the limit of small $\omega$ (large times) diagnoses non-stationarity.
For $\gamma=0$, we observe that $\alpha$ is close to $0$, and the corresponding time series can be regarded as stationary. 
In contrast, for $\gamma=10^2,$ and $10^3$, $\alpha \approx 2$, signalling non-stationary
time series similar to random walks. This non-stationarity implies that the notion of an 
average limit order structure as a function of depth is meaningless.

The fact that the number of limit orders at depth  $\gamma=0$ is stationary 
implies that limit orders placed at the best bid and ask prices can be characterised
by a well-defined distribution and time-dependence structure.
In contrast, the fact that the cumulative numbers of limit orders up to large depths
follow dynamics similar to random walks, and are thus non-stationary, implies 
to a first approximation that orders are put
as random additions, cancellations as well as executions without apparent aggregate
strategy to keep the total order book from accumulating orders as would be
the strategy of a market maker trying to manage and limit the size of its inventory.

Although the time series of the cumulative numbers of limit orders (Eq.(\ref{Cum_V}))
are nonstationary as shown in Fig.\ref{fig2}(b), we checked that the increments
defined by eqs.(\ref{Stationary1}) and (\ref{yjuytkikiotry}) are stationary in a weak sense
(first-order moment and autocovariance do not vary with time). 
This allows us to study the configuration changes of limit orders caused by newly injected, 
canceled, and executed orders at depth $\gamma$ in relation to price changes in a given time interval. 
Let us define the coarse-grained cross correlation function between two variables $A(t)$ and $B(t)$ by
\begin{eqnarray}
Cor(A(t;k\Delta t), B(t;k\Delta t)) &\equiv & \frac{1}{\sigma_A \sigma_B}\average{(A(t;k\Delta t)-\average{A})(B(t;k\Delta t)-\average{B})}~,
\label{Cor1}
\end{eqnarray}
where $\average{A}$ and $\sigma_A$ 
represent the average of variable $A$ and the standard deviation 
(and similarly for variable $B$), and $A(t;k\Delta t)$ is given by
\begin{eqnarray}
A(t;k\Delta t)\equiv \sum_{s=0}^{k}\frac{1}{k\Delta t}A(t+s\Delta t) ~.
\label{CoarseG}
\end{eqnarray}
where $s$ and $k$ are integer numbers. The
coarse graining operation (\ref{CoarseG}) performed over the time scale $k\Delta t$ 
enables one to remove zig-zag and noisy behaviors at very short time scale to extract
the robust large-scale correlation coefficients (\ref{Cor1}) . Note that 
$A(t+k\Delta t;k\Delta t)$ comes after $A(t;k\Delta t)$ as we do not use the 
same data points for estimating the two coarse-grained variables. 

\begin{figure}
\includegraphics[scale=1.2]{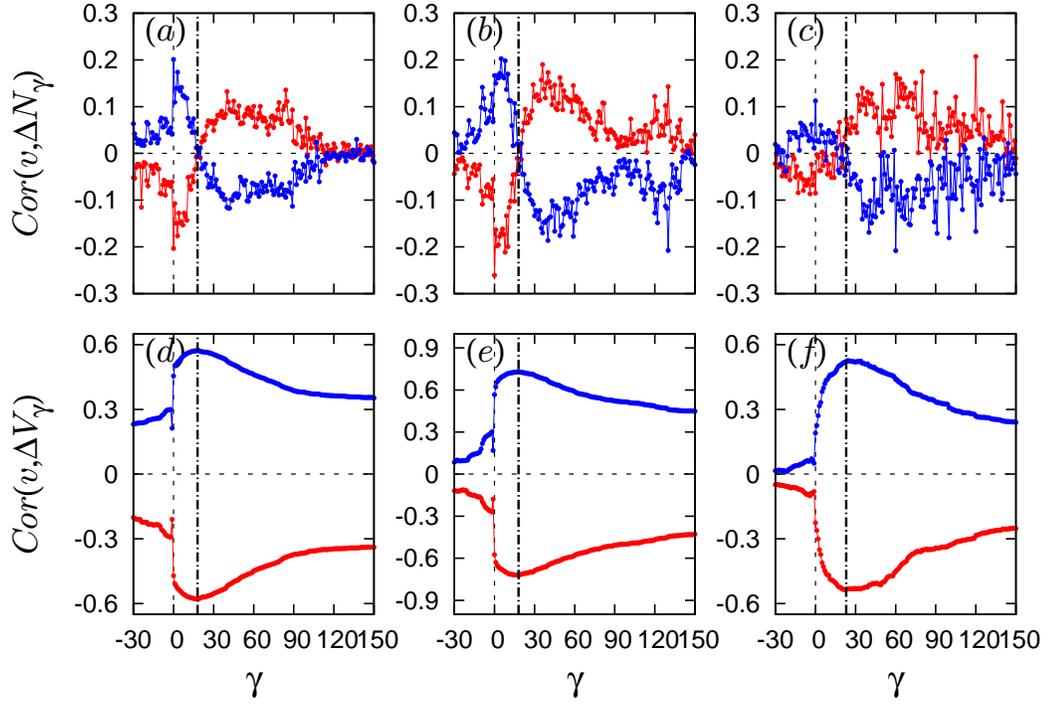}
\caption{
Study of the relationship between the dynamics of the changes with time of the number
of limit orders and the price changes: 
(a) Cross correlation function $Cor(v(t),\Delta N_{\gamma}(\gamma,t;k\Delta t))$ for USD/JPY
as a function of $\gamma$, showing the critical depth $\gamma_c$ defined in the text with the black dotted line; 
(d) Cross correlation function $Cor(v(t),\Delta V_{\gamma}(\gamma,t;k\Delta t))$ for USD/JPY 
as a function of $\gamma$, where the critical depth $\gamma_c$ is indicated with the black dotted line;
Panels (b,e) are the same as panels (a, b) for the EUR/USD exchange rate. 
Panels (c,f) are the same as panels (a, b) for the EUR/JPY exchange rate.
Blue and red circles corresponds to the $-$ and $+$ side.
}
\label{fig3}
\end{figure}

Substituting $v(t;k\Delta t)$ and $\Delta N_{\gamma}(\gamma^j,t;k\Delta t)$ 
for $A(t;k\Delta t)$ and $B(t;k\Delta t)$ in Eq.(\ref{Cor1}), 
the cross correlation coefficients between price changes and the  
changes of the numbers of limit orders at depth $\gamma$ are shown in 
FIG.\ref{fig3}(a,b,c) (a: USD/JPY, b: EUR/USD, c:EUR/JPY). 
Blue and red circles corresponds to the $-$ and $+$ side. 
For these three markets, we observe that configuration changes of limit orders have a common 
functional form of their correlations with price changes. In particular, the cross-correlation
functions change sign at some critical value $\gamma_c$ defined as follows:
\begin{eqnarray}
\gamma_c^+ \equiv Inf\{\gamma>0: Cor(v(t), \Delta N_{\gamma}(\gamma,t))<0, Cor(v(t), \Delta N_{\gamma}(\gamma+1,t))>0 \}
\label{yuyjiitr1} \\
\gamma_c^- \equiv Inf\{\gamma>0: Cor(v(t), \Delta N_{\gamma}(\gamma,t))>0, Cor(v(t), \Delta N_{\gamma}(\gamma+1,t))<0 \}.
\label{yuyjiitr2}
\end{eqnarray}
FIG.\ref{fig3}(a,b,c) show that $\gamma_c^+$ and $\gamma_c^-$ are essentially undistinguishable, 
confirming a symmetric behavior of the structure of the limit order book in the buy $-$ side and sell $+$ side.

Substituting $v(t;k\Delta t)$, $\Delta V_{\gamma}(\gamma^j,t;k\Delta t)$ 
for $A(t;k\Delta t)$ and $B(t;k\Delta t)$ in Eq.(\ref{Cor1}), 
FIG.\ref{fig3}(d,e,f) shows the estimated correlation functions 
between $\Delta V_{\gamma}(\gamma^j,t;k\Delta t)$ and 
$v(t;k\Delta t)$, where the blue and red colour corresponds to the $-$ and $+$ side (d: USD/JPY, e: EUR/USD, f: EUR/JPY). 
One can observe that the maximum of these correlation functions 
$Cor(v(t;k\Delta t), \Delta V_{\gamma}(\gamma^j,t;k\Delta t))$ occurs at the previously defined
critical values $\gamma_c^{+,-}$ (\ref{yuyjiitr1},\ref{yuyjiitr2}). Thus, locating the peaks of 
$Cor(v(t;k\Delta t), \Delta V_{\gamma}(\gamma^j,t;k\Delta t))$
provides a convenient and robust way for the 
numerical estimation of $\gamma_c^{+,-}$ from the data:
\begin{eqnarray}
\gamma_c^j = sup\{\gamma^j>0: |Cor(v(t;k\Delta t), \Delta V_{\gamma}(\gamma^j,t;k\Delta t))|\}. 
\label{tyjujiue}
\end{eqnarray}
The dotted vertical lines in FIG.\ref{fig3}(d,e,f) show the location of 
$\gamma_c$ obtained from (\ref{tyjujiue}): $(\gamma^-_c,\gamma^+_c)=(18,18)$ for USD/JPY, 
$(\gamma^-_c,\gamma^+_c)=(17,18)$ for EUR/USD, 
$(\gamma^-_c,\gamma^+_c)=(22,24)$ for EUR/JPY, 
where $k$ is set equal to $20$.
 
These results quantifying the correlations between price increments and changes
of the number of the limit orders as a function of depth express a collective trend following
behavior of traders. 
As price goes up, they tend to requote their limit orders 
at slightly higher price than the best bid. 

\section{Knudsen number of financial markets}

\subsection{Financial Brownian particle in the layered order book fluid}
 
Previous works have already mentioned an analogy between the order book configuration and evolution
on the one hand and a fluid of interacting particles on the other hand. In particular, Refs.~\citep{Maslov,Bak1997430,Challet2001285,2001cond.mat..1474C} 
note similarities between physics and finance \cite{SornetteRev2014}, namely between particle collisions 
in physics and transactions recorded in the order book. 
In Bak et al.'s view~\citep{Bak1997430}, orders are particles and transactions are collisions. 
In Challet and Stinchcombe's view~\citep{Challet2001285}, orders are particles, 
submission of limit orders are particle depositions, cancellation of limit orders represents evaporation,
and transactions correspond to particle annihilation. 

\begin{figure}
\includegraphics[scale=1]{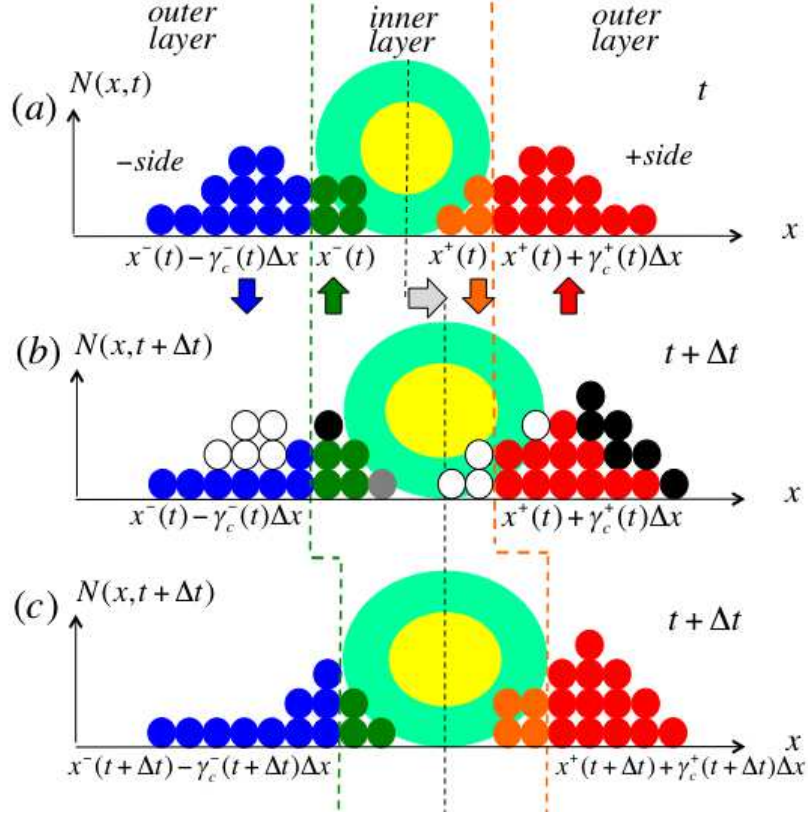}
\caption{
Financial Brownian Particle (FBP) represented by the yellow disk with size $\{x^+-x^-+\Delta x\}/\Delta x$., surrounded
by the fluid of particles corresponding to the different limit orders existing at time $t$. 
(a) A configuration of the FBP and surrounding particles is represented
along the horizontal position $x$ of the order book price and the vertical axis $N(x,t)$
counting the particles (orders).
The vertical dashed line colored with green indicates the position 
$\gamma_c^-\Delta x$ from the best price in the $-$ side. 
Orange one indicates the position $\gamma_c^+\Delta x$ from the best price in the $+$ side. 
The black dotted line shows the position $x(t)$ of the centre of the FBP. 
(b) A configuration change $N(x,t+\Delta t)$ occurring between $t$ and $t+\Delta t$. 
$N(x,t+\Delta t)$ gives the new numbers of limit orders as a function of price $x$.
The newly created particles on the side $\gamma \ge 0$ and $\gamma<0)$ are represented in black 
and grey. The annihilated orders are shown in white.
Upward and downward colored arrows indicate the increase and decrease of the number of 
particles in each range. 
(c) New FBP position and configuration $N(x,t+\Delta t)$ of the surrounding particles
at time $t+\Delta t$ 
Dotted coloured vertical line have moved from their previous positions at time $t$ (panel (a)) 
because of the modification of the best prices, $x^-(t)$ and $x^{+}(t)$. 
Correspondingly, the interaction range is also updated, which revise the colours of the particles. 
}
\label{fig3_2}
\end{figure}

Ref. \citep{PhysRevLett.112.098703} has pushed such qualitative analogies to the quantitative level.
We introduced a description of the dynamics
of the order book of financial markets in terms of a Brownian particle 
embedded in a fluid of incoming, exiting and annihilating particles, as illustrated by Fig.\ref{fig3_2}.
The Financial Brownian Particle (FBP) is represented by the yellow disk, surrounded
by the fluid of particles corresponding to the different limit orders existing at time $t$. 
The vertical dashed line colored with green and orange indicates the position 
$\gamma_c\Delta x$ from the best price in the $-$ and $+$ side. 
The range $ x^- -\gamma_c^-\Delta x \le x  \le x^+ +\gamma_c^+\Delta x$ is called Inner.
The range $x > x^+ +\gamma_c^+\Delta x$, $x < x^- -\gamma_c^-\Delta x$ is called Outer layer. 
The intervals $ x^+ \le x  \le x^+ +\gamma_c^+\Delta x$ and $ x^- -\gamma_c^-\Delta x \le x  \le x^-$ 
are called ``interaction ranges'' within which the FBP interacts with its surrounded particles. 
The yellow circle delineates the space occupied by the Financial Brownian particle, whose size is $\{x^+-x^-\}/\Delta x$. 

Fig.\ref{fig3_2}(b) shows a configuration change occurring between $t$ and $t+\Delta t$.
The newly created particles on the side $\gamma \ge 0$ and $\gamma<0)$ are represented in black and grey. The annihilated orders are shown in white.
Upward and downward colored arrows indicate the increase and decrease of the number of 
particles in each range, which is supported by the statistical analysis presented previously in FIG.\ref{fig3}. 
Fig.\ref{fig3_2}(c) shows the new FBP position and configuration of the surrounding particles,
with in particular the update of the market depth $\gamma$ associated with the changes of
the positions of $x^+$ or $x^-$ from $t$ to $t+\Delta t$.

Within this financial Brownian particle in the layered order book fluid model \citep{PhysRevLett.112.098703},
the positive correlation $Cor(v(t;k\Delta t), \Delta N_{\gamma}(\gamma^j,t;k\Delta t))$ shown 
in FIG.\ref{fig3}(a,b,c) (a: USD/JPY, b: EUR/USD, c:EUR/JPY)
for $\gamma < \gamma_c^-$ (blue curves on the left side of the black dotted line)
is interpreted as resulting from a force that pushes the price towards the $+$ side.
Symmetrically,  the negative correlation $Cor(v(t;k\Delta t), \Delta N_{\gamma}(\gamma^j,t;k\Delta t))$
for $\gamma < \gamma_c^+$ (red curves on the left side of the black dotted line)
is interpreted as resulting from a force that pushes the price towards the $-$ side.
The change of sign of $Cor(v(t;k\Delta t), \Delta N_{\gamma}(\gamma^j,t;k\Delta t))$
when $\gamma$ crosses $\gamma_c^{+,-}$ corresponds to a reversal of the force
exerted by the change of limit orders that plays the role of a drag resistance 
against the direction of the price change.

\subsection{Definition of the mean free path}

Using the above formalism of the FBP, we now relate the variations of the number 
of particles to the motion of the FBP characterised by its velocity $v(t)$.
The change in the number of particles belonging to the 
inner layer that is caused by newly created particles
is noted $c^i(t)(>0)$ and by newly annihilated particles is noted $a^i(t)(>0)$. 
The change $f^j_i(t)$ in the number of particles in the inner layer (hence the index $i$) on the $j$ side is defined as
\begin{eqnarray}
f_i^j(t) & \equiv & \sum_{\gamma=0}^{\gamma_c^j} \Delta N_{\gamma}(\gamma^j,t) \label{trhejuyj}\\
&\equiv& c^j_i(t)-a^j_i(t). \label{ip}
\end{eqnarray}
In Fig.\ref{fig3_2}, $c^-_i(t)$ corresponds
to the particles in the green inner layer. 
$c^+_i(t)$ corresponds to the particles in the orange inner layer. 
These added particles 
are specifically identified by the black colour. 
As depicted in Fig.\ref{fig3_2}, $a^-_i(t)$ and  $a^+_i(t)$ 
corresponds to the particles that are removed among the green and orange particles
in the inner layer. These removed particles are represented in white in the inner circle.
Then, $f_i^j(t)$ is the total flow of particles from $t$ to $t+\Delta t$. 

As shown in Fig.\ref{fig3}(d,e,f), the cross correlation coefficient between 
$f_i^j(t)$ and $v(t)$ takes the largest possible (absolute) value compared 
to correlations between $v(t)$ and changes
in particle numbers that would be counted up to values $\gamma$ different from 
$\gamma_c$. In fact, the variable $f_i(t)$ that has the highest cross correlation coefficients with $v(t)$ is
derived from $f_i^j(t)$, with 
$j=+,-$ (\ref{trhejuyj}) 
as
\begin{eqnarray}
f_i(t)\equiv f_i^-(t)-f_i^+(t).\label{f_i}~
\label{eytjrw}
\end{eqnarray}
Optimizing the coefficients $\alpha$ and $\beta$ of the ansatz $\alpha f_i^-(t) + \beta f_i^+(t)$
to maximise correlation with $v(t)$ gives results that are statistically not significantly
better than expression (\ref{eytjrw}) (which corresponds to $\alpha = - \beta =1$). 

\begin{figure}
\includegraphics[scale=1.2]{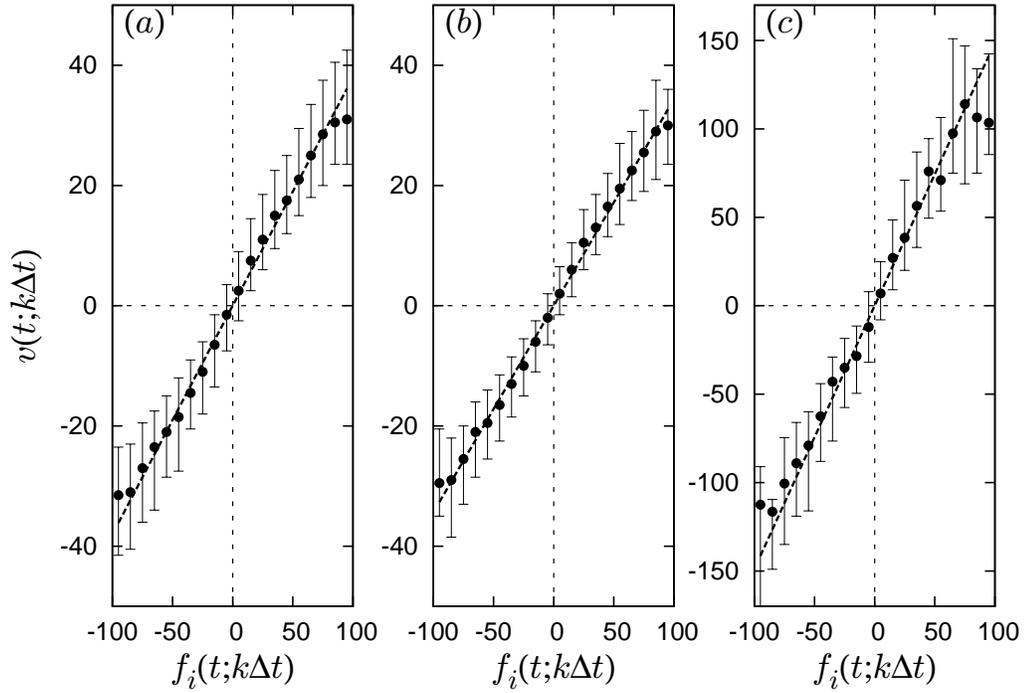}
\caption{
Empirical evidence for Eq.(\ref{eq}):
(a) For USD/JPY, scatter plot of $v(t;k\Delta t)$ as a function of $f_i(t;k\Delta t)$, with $k=20$.
The median is the thick blue dot and the error bars are the inter-quartiles. 
The fitted dotted line is estimated by the least square method.
(b) Same as (a) for EUR/USD. (c) Same as (a) for EUR/JPY.
}
\label{fig3_3}
\end{figure}

Using Eq.(\ref{CoarseG}), we next investigate the relation between the coarse-grained variables 
$f_i(t;k\Delta t)$ and $v(t;k\Delta t)$.
Fig.\ref{fig3_3}(a,b,c) shows the existence of a clear linear relation between these variables, which 
can be expressed as
\begin{eqnarray}
v(t;k\Delta t)=L(t;k\Delta t) f_{i}(t;k\Delta t) + \eta(t;k\Delta t)
\label{eq}
\end{eqnarray}
where $\eta(t;k\Delta t)$ embodies independent noise terms.  Here, we have use $k=20$.
The slope $L(t;k\Delta t)$ of the regression has the meaning of a transport coefficient, 
which is measured in the unit [$\Delta x/\Delta n$]. 

Within the physical picture of a Brownian particle 
embedded in a fluid of incoming, exiting and annihilating particles, $L$ can be interpreted
as the mean free path of the FBP in the order book fluid.
Indeed, in molecular dynamics, the mean free path $L$, collision time $\tau$ and
short-time ``ballistic'' velocity $v$ (between collisions) are related by the formula  \cite{Sone2002}
\begin{equation}
v = {L \over \tau}~.
\label{wrthjytju}
\end{equation}
Since there are $f_{i}(t;k\Delta t)$ collisions in a time interval of duration $k\Delta t$,
as obtained by averaging over a time interval of duration $k\Delta t)$ (\ref{CoarseG}),
the typical collision time (i.e. time between two collisions) is
\begin{equation}
\tau = {k\Delta t \over f_{i}(t;k\Delta t)}~.
\label{wrthjwsytju}
\end{equation}
Putting (\ref{wrthjwsytju}) in (\ref{wrthjytju}) yields expression (\ref{eq}) (without
the residual term), when counting time in units of $k\Delta t$.

Estimating $L(t;k\Delta t)$ from the regressions shown in Fig.\ref{fig3_3}(a,b,c), we obtain
respectively $L(t;k\Delta t) =0.38,0.34,1.49$ for USD/JPY, EUR/USD, EUR/JPY. 
Thus, the mean free path depends on the market.

 As we will show below, financial markets often exhibit asymmetric behavior.
This motivates us to generalize (\ref{eq}) into two distinct linear relationships for the
$+$ and $-$ sides:
\begin{eqnarray}
v^j(t;k\Delta t)&=&L^j(t;k\Delta t) f^j_{i}(t;k\Delta t) + \eta^j(t;k\Delta t)~,\label{eq2}
\end{eqnarray}
where $j$ is either $+$ or $-$.

\subsection{Definition of the Knudsen number $Kn$}

All financial markets are characterised by discrete price increments $\Delta x$ and 
time stamps (corresponding to the minimum precision time of $0.001$ sec for our forex data).
 The question therefore arises whether the standard financial mathematical formulation
 in terms of continuous stochastic processes \cite{Jeanblancetal2009} hold and under what conditions.
 A similar question is often posed in physics concerning the conditions for the application
 of the continuous Navier-Stokes equation of fluid dynamics, given the fact that the
 underlying fluid is made of discrete molecules. In Physics, this question is addressed
 by introducing the Knudsen number $Kn$, defined as the ratio of the mean free path
 of the fluid particles to a characteristic molecular scale. When $Kn$ is sufficiently 
 smaller than $1$, the continuous limit is a good approximation of the dynamics.
 
 Similarly, our model of a financial Brownian particle in the layered order-book fluid
 offers the possibility of defining and estimating a corresponding Knudsen number, which
 characterises a given market. We thus obtain the novel possibility to address quantitatively for the first
 time the question of whether the use of continuous stochastic processes
 to model financial price dynamics is justified.
 
We have introduced the mean free path $L^j$ in expression (\ref{eq2}).
As a characteristic scale, it is natural to use the representative length scale $\gamma_c^j$ characterising the
interaction range, shown in Fig.(\ref{fig3_2}). Then,  the Knudsen number $Kn$ of a given 
financial market, considering possible asymmetric cases, is defined by 
\begin{eqnarray}
Kn^j \sim \frac{L^j}{\gamma_c^j}. \label{def_Kn2}
\end{eqnarray}
A symmetric version of the financial Knudsen number is given by
\begin{eqnarray}
Kn \sim \frac{1}{2} \left( \frac{L^+}{\gamma_c^+} + \frac{L^-}{\gamma_c^-} \right). \label{def_Kn}
\end{eqnarray}

For its empirical determination, we estimate $L(t;k\Delta t)$ by the least square method  
in the time window $[t-(S\cdot k-1)\Delta t, t]$, which contains $S$ points. In turn,
the characteristic scale $\gamma_c$ is found to be stable as a function of time, so that
we can use a single value for the whole period according to the procedure explained
from FIG.\ref{fig3}(d,e,f). Fig.\ref{fig3_3}(a,b,c) showed an average value of the mean free path
$L(t)$, but this does not mean that $L(t)$ is constant. We find in fact that 
$L(t;k\Delta t)$ in shorter periods (involving $S$ data points) depends on time, so 
that $Kn(t;k\Delta t)$ becomes a time-dependent variable.

\subsection{Condition for the validity of the continuous time formalism: $Kn(t;k\Delta t) < \theta_{Kn}$}

\begin{figure}
\includegraphics[scale=1.2]{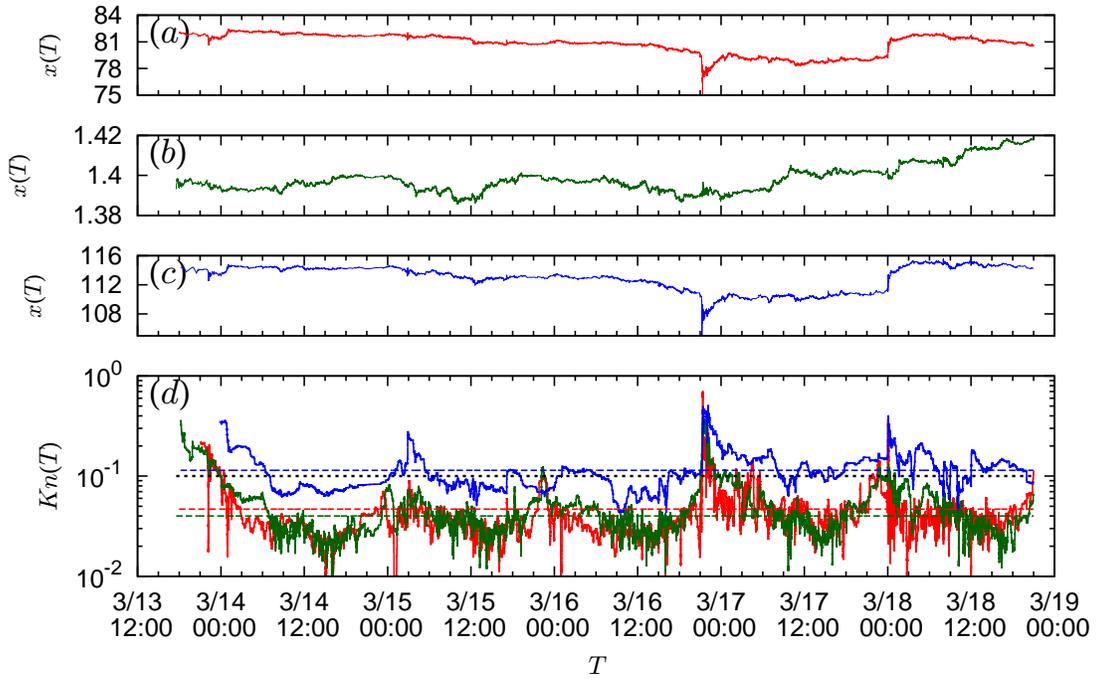}
\caption{
Time evolution of the position $x(T)$ of the FBP
together with the Knudsen number $Kn(T)$, using
$k=4$ and $S=100$ in three markets as a function of calendar time $T$. 
(a) USD/JPY, $x(T)$ (red);
(b) EUR/USD, $x(T)$ (green);
(c) EUR/JPY, $x(T)$ (blue);
(d) $Kn(T)$ for the three currency pairs with the same colour code as
in (a),(b), and (c). The colored dotted lines show the average values of $Kn$.
The black dotted horizontal line shows the level $0.1$.
}
\label{fig4}
\end{figure}

Fig.\ref{fig4} shows the time evolution of the position $x(T)$ of the FBP
together with the Knudsen number $Kn(T)$ estimated as described above using
$k=4$ and $S=100$ in three markets as a function of calendar time $T$. 
We use calendar time to compare markets in which 
transactions do not always occur at the same time.
The three markets are  USD/JPY, EUR/USD and EUR/JPY.
The same color code is used for $x(t)$ and $Kn(t)$ for each currency pair.
The straight horizontal lines in Fig.\ref{fig4}(d) corresponds to the average
level $\average{Kn(T)}$ of the Knudsen number over the time period shown.
The average values of the Knudsen numbers are respectively 
$\{0.046, 0.039, 0.14\}$ for USD/JPY, EUR/USD, EUR/JPY. 
The black dotted horizontal line shows the level $0.1$, which is the 
typical threshold value below which the continuum limit is usually considered to be valid. 

A first conclusion is therefore that the continuous limit seems in general adequate for the 
USD/JPY and EUR/USD currency pairs but is more questionable for the EUR/JPY pair.
For the EUR/JPY pair, this is due to the fact that the position of the FBP (mid-price value) fluctuates
under the influence of a small number of particles (limit orders), which makes
discreteness relevant. 

\begin{figure}
\includegraphics[scale=1.2]{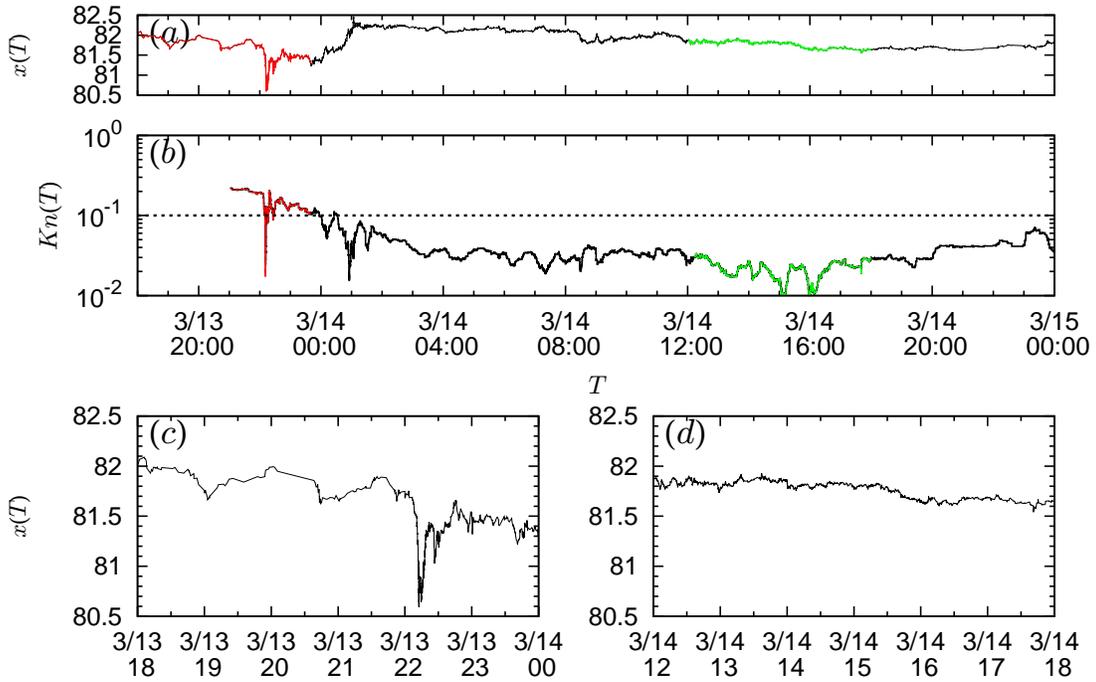}
\caption{
Typical time evolution of the USD/JPY exchange rate and of the corresponding Knudsen number:
(a) Exchange rate time series $x(t)$ (position of the FBP); the red is enlarged in panel (c) and the blue is enlarged in panel (d). 
(b) Estimated Knudsen number corresponding to the time series shown in panel (a).
(c) Enlarged time series of the exchange rate $x(t)$, corresponding to an estimated Knudsen number above $0.1$. 
(d) Enlarged time series of $x(t)$, corresponding to an estimated Knudsen number well below $0.1$. 
}
\label{fig4_2}
\end{figure}

But even for the USD/JPY and EUR/USD pairs, one can observe
that there are times when $Kn(t)$ jumps up, signalling periods when the 
continuous description of the price dynamics becomes incorrect.
Consider Fig.\ref{fig4_2} comparing the time series of the USD/JPY
exchange rate $x(t)$ (position of the FBP) in panel (a) and of its Knudsen number in panel (b).
The time period covers a regime (depicted in red) in which the price is quite volatile
and the Knudsen number is above the $0.1$ threshold, and another regime
in green in which the price has low volatility 
and the Knudsen number is well below the $0.1$ threshold.
Fig.\ref{fig4_2}(c,d) magnify the dynamics of $x(t)$ in these two periods.

\subsection{Relation between mean free path and particle density}

\begin{figure}
\includegraphics[scale=1.2]{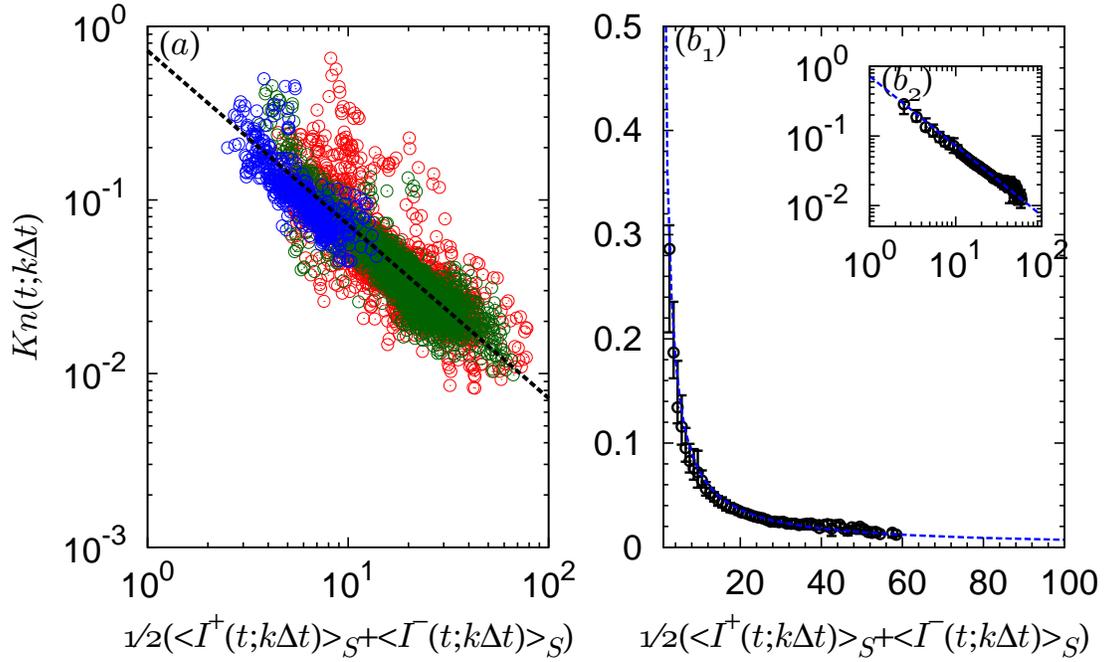}
\caption{
Dependence of the Knudsen number as a function of the number of particles in the inner layer.
(a) Average Knudsen number $Kn(t;k\Delta t)$ in the symmetric case estimated
using expression (\ref{def_Kn}) as a function of the average number $(\average{I^+}_S+\average{I^-}_S)/2$ of 
particles over the two sides, with $S=100$. We use the same color code as in Fig.\ref{fig4} to refer to the 
three pairs of currencies: USD/JPY (red), EUR/USD (green), EUR/JPY (blue).
(b$_1$) Median and quartiles of the data points in (a) together with the best fit with expression (\ref{L_norm})
(dotted blue line) $\kappa=0.72$. 
(b$_2$) log-log plot of (b$_1$) illustrating the inverse relationship (\ref{L_norm}).
}
\label{fig5}
\end{figure}

We now empirically demonstrate that high Knudsen numbers occur when
the number of particles (limit orders) in the interaction range is small, in agreement with the physical
intuition for the breakdown of the continuous approximation.
The total number $I^j(t)$ of particles in the inner layer on the $j$ side is given by
\begin{eqnarray}
I^j(t) &=& \sum_{\gamma'^j=0}^{\gamma^{j}_c} N_{\gamma}(\gamma'^j,t).
\end{eqnarray} 
We average $I^j(t)$  over $S$ data points,
\begin{eqnarray}
\average{I^{j}(t;k\Delta t)}_{S} = \sum_{t'=0}^{S\cdot k -1} \frac{1}{S\cdot k\Delta t}I^{j}(t-t'\Delta t)k\Delta t, 
\end{eqnarray} 
where $t'$ is an integer index. We shall also use this time-average 
over $S$ past time instants for other variables later.
FIG.\ref{fig5}(a) demonstrates that the Knudsen number $Kn$ is on average
inversely proportional to the average number of particles $(\average{I^+}_S+\average{I^-}_S)/2$
over the two sides averaged over $S = 100$ data points, 
\begin{equation}
Kn(t;k\Delta t) \sim 2 \kappa \left( \average{I^+(t;k\Delta t)}_S+\average{I^-(t;k\Delta t)}_S\}\right)^{-1},
\label{L_norm}
\end{equation} 
for the three currency pairs USD/JPY,  EUR/USD and EUR/JPY). In expression (\ref{L_norm}),
we use expression (\ref{def_Kn}) for $Kn(t;k\Delta t)$ in the symmetric case and thus
average the number $(\average{I^+}_S+\average{I^-}_S)/2$ of particles over the two sides.
We estimate $\kappa=0.72$ by the least square method. 
Given a threshold $\theta_{Kn}$ (usually set at $0.1$) above which the continuous approximation is deemed unreliable
and discrete effects start to be important, then the condition for the 
validity of the continuous approximation 
\begin{equation}
Kn(t;k\Delta t) < \theta_{Kn}
\end{equation}
translates into 
\begin{equation}
\frac{1}{2}(\average{I^+(t;k\Delta t)}_S+\average{I^-(t;k\Delta t)}_S) > \kappa/\theta_{Kn} ~,
\end{equation}
i.e., the average number of particles in the inner layer on each side should be at least 
$\kappa/\theta_{Kn} = 7.2$ (for $\theta_{Kn}=0.1$). Summing over both sides, this implies
that at least 14 particles should be present in the inner layer for the continuous approximation
to hold. We note that Ref.~\cite{Cont} found a similar relationship for equities between 
the time-averaged number of limit orders at the best quotes and the price changes. 
Indeed, expression (\ref{eq}) together with (\ref{L_norm}) implies that
$v(t;k\Delta t) \sim \left( \average{I^+(t;k\Delta t)}_S+\average{I^-(t;k\Delta t)}_S\}\right)^{-1}$.

\begin{figure}
\includegraphics[scale=1.2]{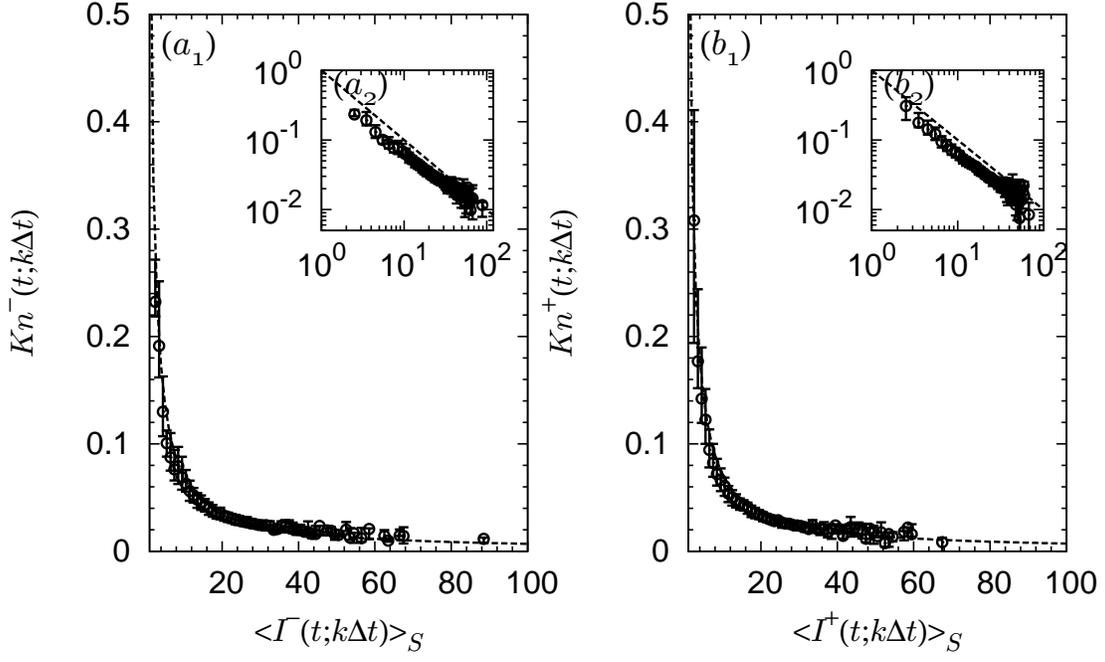}
\caption{
Dependence of the Knudsen number as a function of the number of particles in the inner layer
estimated separately for the $+$ and $-$ sides: 
(a$_1$) Median and quartiles of $Kn^-(t;k\Delta t)$ as a function of $\average{I^-(t;k\Delta t)}_S$ (with $S=100$) for  
the same three currency pairs as in Fig.~\ref{fig5}. The best fit with expression (\ref{L_norm_j})
(dotted line)  gives $\kappa^-=0.69$.
(a$_2$) log-log scale of (a$_1$) illustrating the inverse relationship  (\ref{L_norm_j}).
(b$_1$) same as (a$_1$) for the $+$ side, with $\kappa^+=0.71$. 
(b$_2$) log-log scale of (b$_1$).
}
\label{fig5_2}
\end{figure}

Relation (\ref{L_norm}) can be generalized to include asymmetric cases as follows: 
\begin{eqnarray}
Kn^j(t;k\Delta t) \sim \frac{\kappa^j}{\average{I^j(t;k\Delta t)}_S}. 
\label{L_norm_j}
\end{eqnarray} 
Fig.\ref{fig5_2} shows that Eq.(\ref{L_norm_j}) holds for sides $+$ and $-$
with $\kappa^+ =0.71$ and $\kappa^-= 0.69$, showing an almost symmetry behavior.

Putting together (\ref{def_Kn2}) and (\ref{L_norm_j}) yields
\begin{eqnarray}
L^j(t;k\Delta t) \sim   \kappa^j  \left(\frac{ \average{I^j(t;k\Delta t)}_S }{\gamma_c } \right)^{-1}~. 
\label{L_norm_convert}
\end{eqnarray} 
The variable $\average{I^j(t;k\Delta t)}_S / \gamma_c$ is the ratio of the number of particles in
the inner layer to the size of that layer. Hence, it is the (linear) density $\rho$ of particles in the inner layer.
Thus, equation (\ref{L_norm_convert}) writes that the mean free path defined in equation (\ref{eq})
is inversely proportional to the particle density, which is exactly what is expected
from the kinetic theory of fluids \cite{Sone2002}. Recall the simple argument to derive this
inverse relationship. In a cylinder of length $L$ and cross-section $\sigma$, there are
$\rho L \sigma$ particles. By definition, the mean free path $L$ is such that there is one
particle to collide with, if the cross-section of the collision is $\sigma$. Hence, $L$
is determined by $\rho L \sigma \sim 1$, which leads to $L \sim 1/ \rho \sigma$, which 
has the same form as (\ref{L_norm_convert}), with the identification of 
$\rho$ with $\average{I^j(t;k\Delta t)}_S / \gamma_c$. This confirms further the validity of the 
financial Brownian particle model in the fluid of limit orders. Moreover, the coefficient 
$ \kappa^j$ can be interpreted as the inverse cross-section for the collision of the 
FBP with its surrounding limit order particles.

Putting all this together, Eq.(\ref{eq2}) can be generalised to give the dependence of 
the velocity on the $j$ side as
\begin{eqnarray}
v^j(t;k\Delta t)&=&\frac{\gamma^j_c\cdot \kappa^j}{\average{I^j_i(t;k\Delta t)}_S}\cdot f^j_{i}(t;k\Delta t) + \eta^j(t;k\Delta t)~,
\label{Kn}
\end{eqnarray} 
where $\eta^j(t;k\Delta t)$ is random noise. 
Eq.(\ref{Kn}) stresses that the velocity of the FBP on the $j$ side of the inner layer is inversely proportional to 
the total number of molecules on that side in the inner layer.
This expression captures the result that, when there is an asymmetric change
of the number of particles on both sides of the FBP, 
the total velocity, $\{v^{+}(t;k\Delta t)+v^{-}(t;k\Delta t)\}/2$ will take 
an asymmetric value causing a directed motion. 
In the next section, we document examples of such asymmetric market states.

\section{Asymmetric particle depletion rates, Knudsen numbers criterion and large velocity of the FBP} 

\subsection{Asymmetric particle depletion rates}

The generalisation of expression (\ref{Kn}) (which also makes relation (\ref{eq}) more precise) 
to account for asymmetric particle distributions
on both $+$ and $-$ sides reads on average
\begin{eqnarray}
\average{v(t;k\Delta t)}_S &=& \frac{1}{2}\{\average{v^-(t;k\Delta t)}_{S}+\average{v^+(t;k\Delta t)}_{S}\}\nonumber\\
& \approx & \frac{1}{2} \left(\frac{\gamma^-_c\cdot \kappa^-}{\average{I^-_i(t;k\Delta t)}_S} \average{f^-_{i}(t;k\Delta t)}_S - \frac{\gamma^+_c\cdot \kappa^+}{\average{I^+_i(t;k\Delta t)}_S}  \average{f^+_{i}(t;k\Delta t)}_S \right)~,
\label{time-averaged_velocity}
\end{eqnarray}
where $\gamma_c^j$ and $\kappa^j$ are nearly the same for $j=+$ and $j=-$.
The residual term  $\eta^j$ is averaged out.
  
Let us consider the case when the velocity $v^-$ takes a positive value 
caused by transactions or cancelations that totally deplete the number of particles in the inner layer. 
We want to quantify the average speed needed to wipe out the number of particles in the inner layer. 
For this, we measure the ratio of the change $\average{f_i^j(t;k\Delta t)}_S$ per unit time
of the number of particles in the inner layer to the total average number 
$\average{I^j_i(t;k\Delta t)}_S$ of particles in the inner layer, both on the $j$ side 
\begin{eqnarray}
\lambda^j(t;k\Delta t) \equiv \frac{\average{f_i^j(t;k\Delta t)}_S}{\average{I^j_i(t;k\Delta t)}_S}. \label{lambda}
\end{eqnarray}
Comparing with expressions (\ref{Kn}) and (\ref{time-averaged_velocity}), one can see that $\lambda^j(t;k\Delta t)$
is proportional to the time-averaged velocity on the $j$ side, up to the constant parameters, $\gamma_c^j$ and $\kappa^j$. 

By definition, $\lambda^j(t;k\Delta t)$ is the rate of change of the number of particles in the inner layer
on side $j$. In the continuous limit, this translates into ${1 \over I} {dI \over dt} = \lambda$.
For $\lambda <0$, it thus takes $ \ln 2 / |\lambda|$ ticks to halve the number of particles in the inner layer.
When the number of particles at a certain price level is totally exhausted, the price jumps to the next level.
The typical time scale for this to occur is $\sim 1/ \lambda^j(t;k\Delta t)$.

\begin{figure}
\includegraphics[scale=1.2]{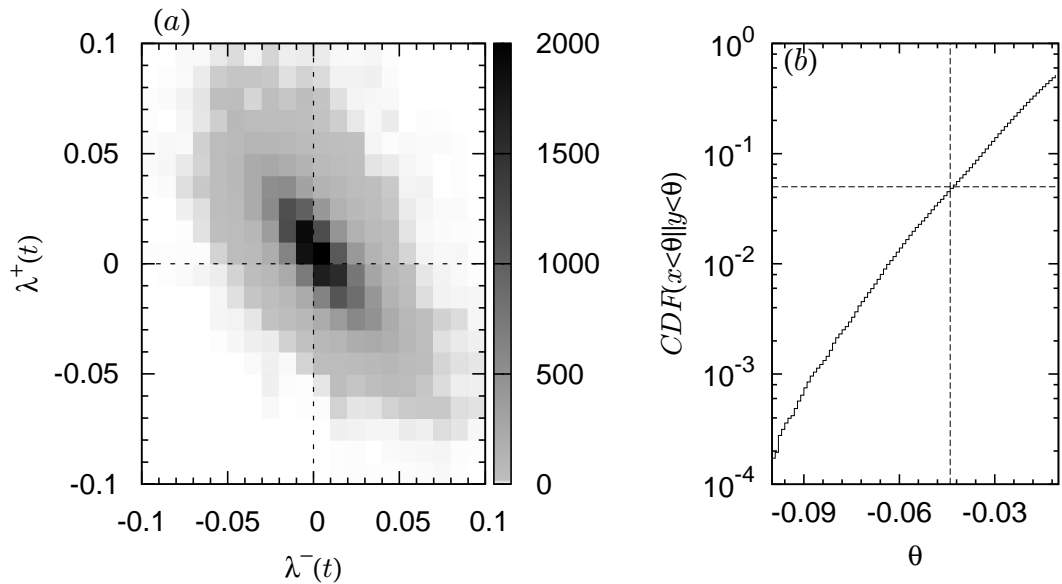}
\caption{
(a) Joint histogram of the rates of change of particle numbers $\lambda^-$ and $\lambda^+$ in the inner layer
exhibiting a negative cross-correlation of $-0.58$.
(b) Cumulative distribution of  $\theta$ defined as the normalised sum over all instances in which 
both $\lambda^-<\theta$ and  $\lambda^+<\theta$ occur simultaneously.
The horizontal dotted line shows the 5\% confidence level, which is reached at $\theta=-0.044$. 
}
\label{fig_ADD}
\end{figure}

Fig.\ref{fig_ADD} shows the two-dimensional joint probability density distribution of 
$\lambda^-$ and $\lambda^+$. The approximate elliptic shape qualifies a bivariable
Gaussian distribution, with a negative cross-correlation $-0.58$ visible by the tilt 
of the long axis of the ellipse along the anti-diagonal. This means that 
$\lambda^-$ and $\lambda^+$ tend to take opposite signs. When particles are piling
up on one side, they tend to disappear on the other side. This embodies 
a transient dominance of buying orders and cancelations. 
The asymmetric directional motion of the FBP often occurs due to an imbalance 
of the rate of change between the two sides, which is mirrored by a corresponding
asymmetry of the changes of the Knudsen numbers on both sides. 

Since large asymmetric depletion rates (imbalance of $\lambda^-$ and $\lambda^+$)
are often leading to strong directed price moves, it is useful to quantify how
frequently this occurs. As a benchmark, Fig.\ref{fig_ADD}(b) shows the cumulative distribution of 
$\theta$ defined as the sum over all instances in which 
both $\lambda^-<\theta$ and  $\lambda^+<\theta$ occur simultaneously.
We find that, in 5\% the time series, $\lambda^-$ and $\lambda^+$ are smaller
than $\theta_{0.05} = -0.044$. At these rates, the number of particles
decays by a factor $2$ in approximately $6.8$ ticks ($10.9$ seconds on average). 
We are however mostly interested in the asymmetric cases when either 
$\lambda^-<\theta$ or $\lambda^+<\theta$ but not both together, expressing a strong 
asymmetric particle depletion on one side. 

\subsection{Joint conditions on asymmetric particle depletion rates and Knudsen numbers to detect
abnormal market regimes}

\begin{figure}
\includegraphics[scale=1.2]{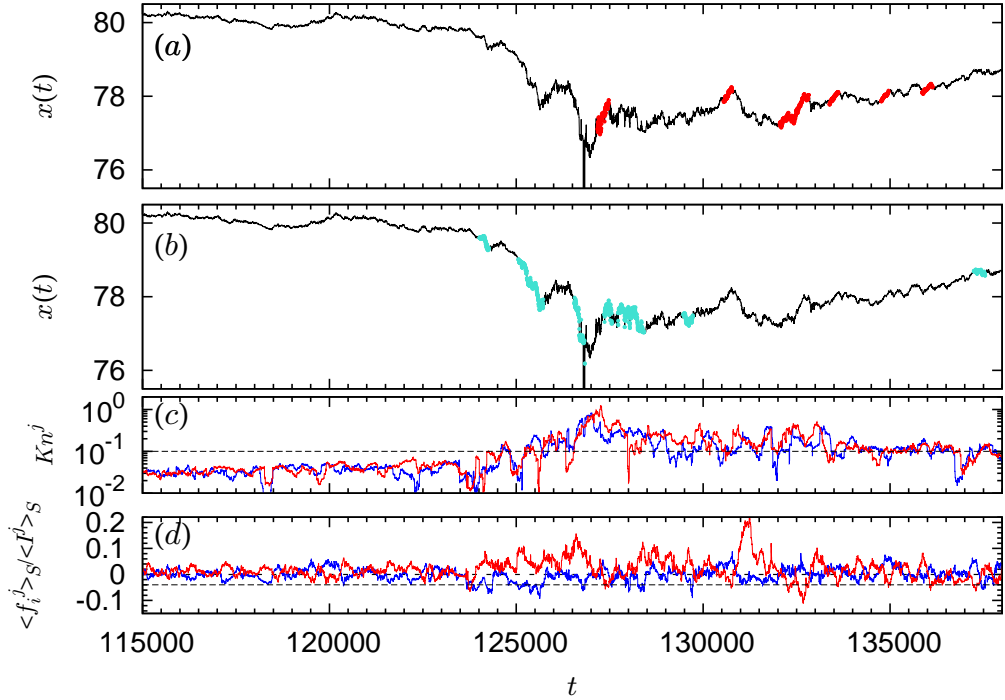}
\caption{
Time dynamics of Knudsen numbers and rates of the change of particle numbers in the inner layer around a flash crash
for the USD/JPY exchange rate on March 16, 2011 (GMT+0). 
On the horizontal axis, time is in multiple units of tick time.
(a) Locations in red  of the USD/JPY exchange rate  when $\lambda^+ < \theta_{0.05}= -0.044$ 
and simultaneously $Kn^+$ is larger than $\theta_{Kn} =0.1$.
(b) Locations in cyan  of the USD/JPY exchange rate  when $\lambda^- < \theta_{0.05}= -0.044$ 
and simultaneously $Kn^-$ is larger than $\theta_{Kn} =0.1$.
For given values at time $t$, the evaluation of these variables is performed in 
the corresponding time interval $[t,t-(S\cdot k-1)\Delta t]$, with $k=2$ and $S=100$. 
(c) Time dependence of the asymmetric Knudsen numbers
$Kn^+(t;k\Delta t)$ (red)  and $Kn^-(t;k\Delta t)$ (blue) 
with the black horizontal dotted line indicating the threshold level $\theta_{Kn}=0.1$.
(d) Corresponding time dependence of 
the rates $\lambda^+(t)$ (red) and  $\lambda^-(t)$ (blue) of particle changes on each side 
with the black horizontal dotted line indicating the threshold level $\theta_{0.05}=-0.044$.
}
\label{fig6_1}
\end{figure}

Owing to the negative correlation coefficient 
documented in Fig.\ref{fig_ADD}(a), we expect more often the depletion rate
to be high on one side while low or average on the other side. 
Fig.\ref{fig6_1}(a) illustrates this point by showing the locations in red in panel (a)
of the USD/JPY exchange rate when $\lambda^+ < \theta_{0.05}= -0.044$. 
Fig.\ref{fig6_1}(b) illustrates this point as well as in cyan in panel (b) when $\lambda^- <  \theta_{0.05}=-0.044$ 
and simultaneously $Kn^+$ (resp. $Kn^-$) is larger than $\theta_{Kn} =0.1$.
For this analysis, the above conditions are deemed valid at time $t$ for an analysis in 
the corresponding time interval $[t,t-(S\cdot k-1)\Delta t]$, with $k=2$ and $S=100$. 
Fig.\ref{fig6_1}(c) shows the time dependence of the asymmetric Knudsen numbers
$Kn^+$ and $Kn^-$ with the black horizontal dotted line indicating the threshold level $\theta_{Kn}=0.1$.
Fig.\ref{fig6_1}(d) presents the corresponding time dependence of 
the rates $\lambda^+(t)$ and  $\lambda^-(t)$ of particle changes on each side.

Around $t=125,000$, the USD/JPY exchange rate dropped violently.
This fall was preceded, accompanied and followed by a very large depletion rate 
on the $-$ side ($\lambda_- <  -0.044$) as shown by the presence of 
the cyan line in Fig.\ref{fig6_1}(b) and by a large increase of both Knudsen numbers  
$Kn^+$ and $Kn^-$ as shown in Fig.\ref{fig6_1}(c). This shows that the 
number of particles in the inner layer decreased due to the increase of transactions and cancelations. 
And $\lambda^-$ (in cyan) became negative, and often below the threshold $\theta_{0.05}=-0.044$, 
while $\lambda^+$ (in red) remained positive, indicating that a strong imbalance in the 
depletion speed of the number of the particles in the inner layer occurred in favour of the $-$ side.
After $t=127,000$, the USD/JPY exchange rate rebounded, with an strong increase
in the depletion rate on the $+$ side ($\lambda_+ <  -0.044$) as shown by the presence of 
the red line in Fig.\ref{fig6_1}(b) around $t=127,000$. 

Fig.\ref{fig7_1} presents the particle configurations during the abnormal times detected in Fig.\ref{fig6_1}. 
Fig.\ref{fig7_1}(a) shows the whole time series of the USD/JPY exchange rate $x(t)$ (black) 
together with the flash crash in red. 
Fig.\ref{fig7_1}(b) plots the time series $x^{-}(t)$ (blue) and $x^{+}(t)$ (red) 
in the time interval represented in red in panel (a). 
Fig.\ref{fig7_1}(c$_1$),(c$_2$) and (c$_3$) show the profiles of the cumulative number of particles
at the best price at the three points indicated in Fig.\ref{fig7_1}(b)
and located respectively at times $t=124000$ (c$_1$) , $t=126000$ (c$_2$), and $t=126803$ (c$_3$).
We show respectively $V_{\gamma}(\gamma^j,t)$ 
in blue for $j=-$ and in red for $j=+$. 
Fig.\ref{fig7_1}(d) shows the cumulative number of particles 
$V_{\gamma}(\gamma^-,t)$ (blue) and $V_{\gamma}(\gamma^+,t)$ (red)
as a function of $\gamma$, in the limit order book. 
 
At the time $t=124000$ (c$_1$) when the exchange rate begins to drop, 
the pile of particles in front of the FBP exerted a resistance against the downward move.
 During the period when the exchange rate was dropping, and focusing on the specifc time $t=126000$ (c$_2$), 
the number of particles on the $-$ side decreased to about 1/8th 
the number found at time $t=124000$ (c$_1$), reflecting the downward price motion towards the $-$ side
that caused a huge annihilation of the particles piled in the $-$ side. 
Finally, at time $t=126803$ (c$_3$), the number of particles on the $-$ side totally vanished.
In other words, there were no traders on the $-$ side. 
The total number of molecules on the $-$ and $+$ side shown in Fig.\ref{fig7_1}(d) as a function of 
the depth in the limit order book indicates that the profile of the number of particles 
on the $-$ side in log-scale 
remained approximately the same during the duration of the flash crash and its rebound.
This confirms that the changes in the number of particles strongly contribute 
to the price formation during the crash. 
Theoretically, the Knudsen number on the $-$ side reaches infinity at the bottom $t=126803$ (c$_3$).
This is a very rare situation when there is no buyer of dollar in the market and no one can sell dollar any more. 
This halt of the market function occurred in the middle of a working hour. 
The above analysis shows that high values of the Knudsen numbers and the 
development of asymmetric properties of the change of particle numbers on each side 
of the inner layer can detect this peculiar event at its early stage. 

\section{Conclusion}

\begin{figure}
\includegraphics[scale=1.2]{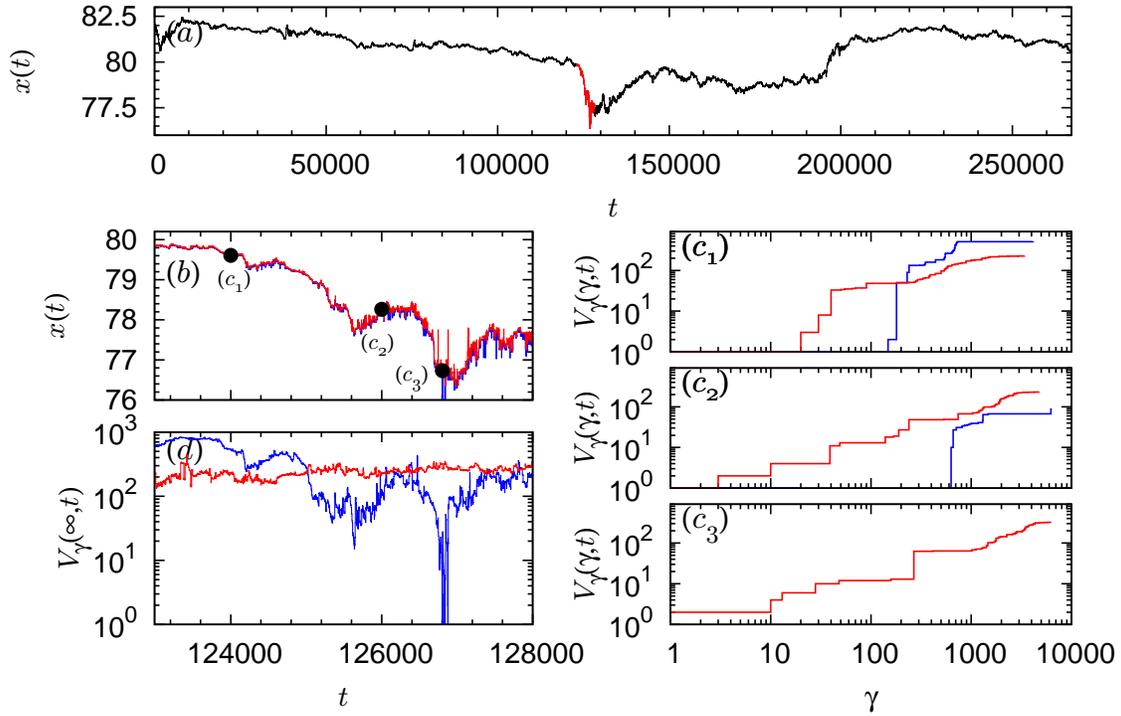}
\caption{
Particle configurations during the abnormal times detected in Fig.\ref{fig6_1}. 
(a) Whole time series of the USD/JPY exchange rate $x(t)$ (black) 
together with the flash crash in red. 
(b) Time series $x^{-}(t)$ (blue) and $x^{+}(t)$ (red) 
in the time interval represented in red in panel (a). 
(c$_1$),(c$_2$) and (c$_3$) show the profiles of the cumulative number of particles
at the best price at the three points indicated in Fig.\ref{fig7_1}(b)
and located respectively at times $t=124000$ (c$_1$) , $t=126000$ (c$_2$), and $t=126803$ (c$_3$).
$V_{\gamma}(\gamma^-,t)$ is shown in blue and 
 $V_{\gamma}(\gamma^+,t)$ is shown in red.
(d) Cumulative number of particles $V_{\gamma}(\gamma^-,t)$ (blue) 
and $V_{\gamma}(\gamma^+,t)$ (red)
in logarithmic scale as a function of $\gamma$ in logarithmic scale in the limit order book.
}
\label{fig7_1}
\end{figure}

Starting from the description of the dynamics of the order book of financial markets 
in terms of a Brownian particle embedded in a fluid of incoming, exiting and annihilating particles
\citep{PhysRevLett.112.098703}, we have generalised this model by presenting
a model of the velocity on each side (buy and sell) independently. The improved model
builds on the time-averaged number $\average{I_i}$ of particles in the inner layer and 
its change $f_i$ per unit time (the index $i$ stands for ``inner layer''). The inner layer
has been defined by specific robust properties of the layering of the particles as
a function of the order book depth,
revealed by the correlations between price velocity and change in the number of particles.
This allowed us to introduce the Knudsen number of the financial Brownian particle (FBP) motion
and its asymmetric version (on the buy and sell sides). Not being considered previously,
the asymmetric Knudsen numbers are crucial in finance in order to detect asymmetric price changes.
We even found rare regimes when no particle stands on one side, so that the corresponding
asymmetric Knudsen number is in principle infinite, as shown in Fig.\ref{fig7_1}. Such situation
is associated with an extraordinary market regime, such as a flash crash, at which
the market stops functioning.

By measuring the Knudsen numbers, we have shown that it is possible to characterise
the conditions for the market dynamics to be correctly described by a continuous stochastic
process.  Even though the continuum formulation is usually applied without 
questions for large liquid markets such as the USD/JPY and EUR/USD exchange rates,
we have shown that there are regimes when the Knudsen numbers are so high that 
discrete particle effects dominate, such as during market stresses and crashes.
Moreover, we found that the EUR/JPY market operates most of the time at 
rather large Knudsen numbers, so that the continuous formulation should be amended.
 
We also found that the Knudsen number is 
inversely proportional to the number of particles (limit orders) in the inner layer. 
This confirms the condition for the breakdown of the continuous formulation, i.e. when
the number of particles is too small for a local average behavior to be defined.
Indeed, small numbers of particles translate in large Knudsen numbers, i.e., 
the mean free path is larger than the typical particle sizes.
If observable, the number of limit orders in the inner layer can be a good indicator 
of the stability and directional movement of financial markets. 

Finally, we documented an imbalance of the depletion rates of the number of particles 
on the buy and sell sides, with the clear occurrence of high Knudsen numbers associated with
violent directional price changes. This indicator is more useful than the 
popular symmetric volatility measure. With these metrics, we could 
detect the direction of the price motion at the early stage while 
volatility is blind to the price direction. 

\bibliographystyle{apsrev4-1}
\bibliography{prl3} 

\end{document}